\documentstyle[amsfonts,amsbsy,12pt,dina4]{article}
\def\fmref#1{(\ref{#1})}

\makeatletter
\def\foot{\small\sl GSI-Preprint 94-13, subm. to Nucl.Phys. {\bf A}}
\begin{document}
\title{A Transport Equation for\\
       Quantum Fields with\\
       Continuous Mass Spectrum 
       \thanks{Work supported by GSI}}
\author{P.A.Henning\thanks{E-mail address: P.Henning@gsi.de}\\
        Institut f\"ur Kernphysik der TH Darmstadt and\\[1mm]
        GSI, P.O.Box 110552, D-64220 Darmstadt, Germany}
\date{\small\sl\today\\ \foot}
\maketitle
\begin{abstract}
Within a relativistic real-time Green's function formalism,
a quantum transport equation for the phase-space
distribution function is derived without a quasi-particle approximation.
Dissipation is due to a nonzero spectral width, and can be 
separated into time-local and memory effects.
\\[2cm]
\noindent
Dedicated to Prof. Hiroomi Umezawa on occasion
of his 70th birthday.
\end{abstract}
\makeatother
\clearpage
\section{Introduction}
Currently one of the main interests of nuclear physics is the
investigation of hot, compressed nuclear matter. One
of the most widely used theoretical tools for this
investigation is the simulation of nuclear collisions by
numerical solution of transport equations,
like e.g. the Vlasov-Uehling-Uhlenbeck (VUU) equation \cite{GCM93}.
This equation and its many relatives have their 
origin in the kinetic theory of molecular gases. Today however
they are obtained from relativistic
quantum field theory, i.e., by processing the Schwinger-Dyson
equation for propagators of the interacting quantum fields
with certain semi-classical approximations \cite{D84a,RS86,MD90}.

It is in principle possible to compare the solution of such
transport equations to the solution of the full Schwinger-Dyson
equation, but to common knowledge such an attempt has not been made but once
\cite{D84a}. The result was a significant difference in the time
evolution, the full quantum system relaxing somewhat slower than
the ''molecular'' transport solution. It has been made plausible that such
an effect is due to ``off-shell'' propagation of particles
\cite{M86,NBH93}, which gives rise to memory effects in the
hot and dense medium \cite{WR93,R94}. A derivation of such memory effects
within the framework of quantum field theory has not been presented
so far.

It is the purpose of the present paper to bridge this gap between
the full Schwinger-Dyson equation and the traditional ``molecular'' 
kinetic picture. 

To this end, one has to go beyond the semi-classical (naive) particle picture
when processing the integral equation for the propagator.
This naive particle picture, traditionally labeled 
``quasi-particle'' approximation, assumes the existence of infinitely
long living elementary excitations even in statistical systems.

Experimentally, there are indications that this is not very well justified:
Spectral functions of hadrons in the medium presumably determine the
spectrum of emitted lepton pairs \cite{HFN93}, and ``momentum
dependent forces'' are thought to be important for the
so-called equation of state in hadronic matter.
In relativistic systems such a momentum dependence
is equivalent to a nonzero imaginary part of the self
energy, and hence again to a nontrivial hadronic spectral
function.  Hence one may adopt the viewpoint, that the extraction of 
{\em quantitative} information from nuclear collisions requires to go
beyond the quasi-particle level. 

Also theoretically the naive particle picture is not a good startpoint:
Consider the breaking of Lorentz invariance 
in matter or temperature states \cite{BS75}. It leads to the 
Narnhofer-Thirring theorem stating the impossibility of a
perturbation theory with quasi-particles at finite temperature
\cite{NRT83}. In other words, the entities to use for a
perturbative description in statistical systems must have a
continuous mass spectrum  -- they {\em cannot\/} 
be quasi-particles. How one may formulate a finite temperature perturbation
theory for quantum fields with continuous mass spectrum has
been shown not too long ago \cite{L88}. Recently we have extended
this picture to time dependent  
\cite{NUY92,hu92} as well as spatially inhomogeneous \cite{h93trans}
non-equilibrium situations.

Here, these recent developments will be extended to a general
non-equilibrium system. The basic assumption for this work is, 
that nuclear phenomena can be described successfully using methods 
from relativistic quantum
field theory. First, to ensure compatibility with 
existing literature, the Schwinger-Dyson equation for
the full propagator of an interacting fermionic quantum field
is processed according to standard procedures. Next, the
equation for the spectral function is solved consistently.
The final step of the present paper is a study of the equation
resulting from insertion of the spectral function into the transport
equation.
\section{Matrix valued Schwinger-Dyson equation}
As has been pointed out by various authors,
the description of dynamical (time dependent) quantum phenomena
in a statistical ensemble necessitates a formalism with a doubled
Hilbert space \cite{D84a,RS86}.  For our purpose the
relevant content of this formalism is, that its two-point Green
functions are 2$\times$2 matrix-valued. We leave it to the reader
to chose either the conventional Schwinger-Keldysh,
or Closed-Time Path (CTP) Green
function formalism \cite{MD90,SKM}, or the technically
simpler method of Thermo Field Dynamics (TFD) \cite{hu92,Ubook}.

Although this choice does not affect the following
considerations, it has to be stated that the principles of TFD
are in accordance with a fundamental requirement: The existence
of two mutually commuting representations of the field operator
algebra. This requirement is not fulfilled by the more conventional CTP (or
Schwinger-Keldysh) formalism, i.e., CTP may lead to mathematical
pitfalls. The recently discussed factorization problem seems to point
in this direction \cite{E93}.

Within this matrix formulation, we consider
the Schwinger-Dyson equation for the full fermion propagator
of a given model
\begin{equation}\label{k1}
           S =       S_0    +      S_0 \odot \Sigma \odot S
\;.\end{equation}
Here $S_0$ is the free and $S$ the full two-point Green function
of the fermion field, $\Sigma$ is the full self energy
and the generalized product of these is to be understood as
a matrix product (thermal and spinor indices) and an
integration (each of the matrices is a function of two space
coordinates):
\begin{equation}\label{ovu}
\Sigma^{ij}_{xy}\odot G^{jk}_{yz} = \sum\limits_j\int\!\!d^4y\,
              \Sigma^{ij}_{xy} G^{jk}_{yz}
\;.\end{equation}
Throughout this paper we use the convention to write
space-time and momentum variables as
lower indices, e.g. $\Sigma_{xy}\equiv \Sigma(x,y)$.

In the CTP formulation as well as in the $\alpha=1$ parameterization
of TFD \cite{hu92}, the matrix elements of $S$ obey
\begin{equation}\label{sme}
S^{11}+S^{22}=S^{12}+S^{21}
\;,\end{equation}
and a similar relation holds for the free propagator $S_0$. It follows that
a linear relation exists also among the components
of the self energy:
\begin{equation}\label{sse}
\Sigma^{11}+\Sigma^{22}=-\Sigma^{12}-\Sigma^{21}
\;.\end{equation}
Thus, the four components of the Schwinger-Dyson equation are
not independent, the matrix equation can be simplified by a linear 
transformation. This linear transformation, which one may
conveniently express as a matrix transformation \cite{RS86,hu92},
has a physical interpretation {\em only} in the TFD formalism: It
is the Bogoliubov transformation defining stable {\em statistical}
quasi-particle states \cite{Ubook,hu92a}, not to be confused
with the traditional quasi-particle approximation. 

In view of this fact, the transformation matrices ${\cal B}$ will be written 
in the form 
\begin{equation}\label{lc}
{\cal B}(n) =
\left(\array{lr}(1 - n) &\; -n\\
                1     & 1\endarray\right)
\;,\end{equation}
depending on one parameter only. In appendix \ref{ling}, some useful
properties of such matrices are listed. Due to the linear relations \fmref{sme}
and \fmref{sse} {\em any} pair of Bogoliubov matrices ${\cal B}(n_1)$,
${\cal B}(n_2)$  gives
\begin{equation}\label{qptp}
  {\cal B}(n_1)\,\tau_3\,S_0\odot\Sigma\odot S\,({\cal B}(n_2))^{-1}
  = \left({\array{lr}  S_0^R\odot\Sigma^R\odot S^R & \mbox{something} \\
                       & S_0^A\odot\Sigma^A\odot S^A \endarray}\right)
\;.\end{equation}
Here, $\tau_3 = \mbox{diag}(1,-1)$,
$\Sigma^{R,A}$ are the retarded and advanced full self energy function,
and $S^{R,A}$ are the retarded and advanced full propagator 
(similarly for $S_0$)
\begin{eqnarray}\label{sra}\nonumber
\Sigma^R = \Sigma^{11}+\Sigma^{12}\;,\;\;\;\;
&\Sigma^A = \Sigma^{11}+\Sigma^{21}\\ 
S^R =  S^{11}-S^{12}\;,\;\;\;\; &S^A = S^{11}-S^{21}
\;.\end{eqnarray}
Together with eqn. \fmref{dig} follows, that
the diagonal elements of the transformed equation therefore
are {\em retarded\/} and {\em advanced\/} Schwinger-Dyson equation.
The off-diagonal element is a {\em transport equation\/}.

We proceed by applying Dirac differential operators
to the diagonal parts of the simplified equation. They are acting
on the free Green function $S_0$ of a fermion field with
mass $M$ according to
\begin{equation}\label{s0sa}\nonumber
\widehat{S}^{-1}_0  = \; {\mathrm i} \partial\!\!\!\mbox{\large/}_x - M
     \Longrightarrow \; \widehat{S}^{-1}_0\,
      S^{A,R}_{0,xy}\;=\;\delta_{xy}
\;,\end{equation}
where 
$\partial\!\!\!\mbox{\large/}_x=\partial^\mu_x\gamma_\mu$
and $\delta_{xy}=\delta^4(x-y)$. The action of these differential
operators on the off-diagonal elements $S_0^{12}$ and $S_0^{21}$ of the
free propagator gives zero. For the retarded and advanced
Schwinger-Dyson equation this yields
\begin{equation}\label{k3}
\widehat{S}^{-1}_0 S^{R,A}_{xy}  =  \delta_{xy} + \Sigma^{R,A}_{xz}
   \odot S^{R,A}_{zy}
\;.\end{equation}
To get rid of the dependence on coordinate differences
present already in space-time homogeneous systems,
one performs a Fourier transform of these equations into
the mixed (or Wigner) representation, according to
\begin{equation} \label{wigdef}
\tilde{\Sigma}_{XP} = \int\!\!d^4(x-y) \;
  \exp\left({\mathrm i} P_\mu (x-y)^\mu\right)\; \Sigma_{xy}
\;.\end{equation}
Here we have set $X=(x+y)/2$, and the ''$\sim$'' will be
dropped from now on. The differential operators (inverse free
propagators) occurring in the equations \fmref{k3} then take the
form
\begin{equation}\label{smix}
\widehat{S}_0^{-1}        = \;
P\!\!\!\!\mbox{\large/} -M +\frac{{\mathrm i}}{2} 
\partial\!\!\!\mbox{\large/}_X
     \Longrightarrow  \; \widehat{S}_0^{-1} S^{A,R}_{0,XP}\;=\;1
\;.\end{equation}
The Wigner transform of the convolution $\Sigma\odot G$ as defined 
in \fmref{ovu} is a nontrivial step: Formally it may be expressed as 
a gradient expansion
\begin{equation}\label{gex}
 \int\!\!d^4(x-y) \;
  \exp\left({\mathrm i} P_\mu (x-y)^\mu\right)\; \Sigma_{xz}\odot G_{zy}
 = \exp\left(-{\mathrm i}\Diamond\right)\,\tilde{\Sigma}_{XP} \,
  \tilde{G}_{XP}
\;.\end{equation}
$\Diamond$ is a 2nd order differential operator acting on both
functions appearing behind it,
\begin{equation}
\Diamond    =  \frac{1}{2}\left(\partial^\Sigma_X\partial^G_P-
                            \partial^\Sigma_P\partial^G_X\right)
\;.\end{equation}
We have indexed the derivatives to avoid confusion on which factor
they act. The index $\Sigma$ stands for real or imaginary part 
of the self energy function, and the index $G$ stands for action on 
real and imaginary part of the propagator. Explicitly, the first-order term is
\begin{equation}\label{dia1}
\Diamond \,\Sigma\, G = \frac{1}{2}\left(
   \frac{\partial \Sigma}{\partial X_\mu}
   \frac{\partial G}{\partial P^\mu}
  -\frac{\partial \Sigma}{\partial P_\mu}
   \frac{\partial G}{\partial X^\mu} \right)
\;.\end{equation}
Obviously, this is the Poisson bracket of the two quantities
$G$ and $\Sigma$. The exponential of $\Diamond$ is a 
differential operator of infinite order, which formally can be 
split into real and imaginary part as 
$\exp\left(-{\mathrm i}\Diamond\right) = 
 \cos\Diamond-{\mathrm i}\sin\Diamond$.

It is not completely clear how to justify the gradient expansion
for arbitrary self energy functions, to infinite or to finite order. 
One may however adopt the following viewpoint on the gradient expansion: It
is the expansion of the expectation value of a composite operator
into a series of terms with increasing non-locality. Such an expansion
is always associated with a length scale, and therefore we have to assume
that the non-equilibrium system is studied only on length
and time scales for which the gradient expansion is rapidly convergent.

Since each operator $\Diamond$ needs a factor $\hbar$ to render it
dimensionless, the gradient expansion is also an expansion in powers
of $\hbar$. Truncating it therefore means, that quantum effects 
occurring on very short length or time scales must be absorbed into
the propagator and the self energy function on larger scales.
Indeed, to use a fashionable expression for this viewpoint: When taking the
gradient expansion seriously, we are considering
an {\em effective field theory\/}. This requires
that the fields we are considering have a 
continuous mass spectrum \cite{W67}.

The propagator and the self energy function in mixed representation
are split into real and imaginary part according to
\begin{eqnarray}\nonumber
S^{R,A}_{XP} &=& G_{XP} \mp {\mathrm i} \pi {\cal A}_{XP} \\
\Sigma^{R,A}_{XP} &=&\Sigma^\delta_X +
  \mbox{Re}{\Sigma}_{XP} \mp {\mathrm i}\pi \Gamma_{XP}
\;.\end{eqnarray}
Each of these functions is assumed to be real, and they 
are spinor-valued. $\Sigma^\delta$ is the part of the
self energy function which is local in space and time, 
which for a relativistic model implies that it is the
Hartree part of $\Sigma$.

It is tempting to call ${\cal A}$ the spectral
function of the fermions. However, it cannot be
guaranteed that the full fermion propagator in non-equilibrium states
has a spectral decomposition \cite{Ubook}. Therefore, this
interpretation has to be treated with caution for the moment.

The gradient expansion and the decomposition of the 
propagator and self energy function are inserted into eqns. \fmref{k3},
leading to 
\begin{eqnarray}\label{k7a}\nonumber
\left(P\!\!\!\!\mbox{\large/} -M +\frac{{\mathrm i}}{2}
  \partial\!\!\!\mbox{\large/}_X - \exp(-{\mathrm i}
  \Diamond)\,\Sigma^\delta_X
  \right)  {\cal A}_{XP}& =&  
\exp(-{\mathrm i}\Diamond)  
 \left[ \mbox{Re}{\Sigma}_{XP} \, {\cal A}_{XP}
       \vphantom{\int} + \Gamma_{XP} \, G_{XP}\right]\\ \nonumber
\left(P\!\!\!\!\mbox{\large/} -M +\frac{{\mathrm i}}{2}
  \partial\!\!\!\mbox{\large/}_X - \exp(-{\mathrm i}\Diamond)\,\Sigma^\delta_X
  \right)  G_{XP}  &=&1\;+\\ \label{k7b}
&&\exp(-{\mathrm i}\Diamond) 
   \left[ \mbox{Re}{\Sigma}_{XP} \, G_{XP}\vphantom{\int}
        -\pi^2 \Gamma_{XP} \, {\cal A}_{XP}\right]
\;.\end{eqnarray}
The above equations contain a mixture of real and imaginary 
part, which can be separated easily. However, they are 
not independent in another sense -- which can be seen when
performing the separation and taking the trace over
the Dirac indices. This gives two pairs of equations:
\begin{eqnarray}\label{k8a}\nonumber
\mbox{Tr}\left[\left(
  \partial\!\!\!\mbox{\large/}_X + 2 \sin\Diamond\;\Sigma^\delta_X
  \right)  {\cal A}_{XP}\right]& =&
  -2 \sin\Diamond\; \mbox{Tr}\left[ \mbox{Re}{\Sigma}_{XP} \, {\cal A}_{XP}
   \vphantom{P\!\!\!\!\mbox{\large/}}
                         + \Gamma_{XP} \, G_{XP}\right]\\ 
\mbox{Tr}\left[\left(  
  \partial\!\!\!\mbox{\large/}_X + 2 \sin\Diamond\;\Sigma^\delta_X
  \right)  G_{XP}\right]& =& 
  -2 \sin\Diamond\; \mbox{Tr}\left[ \mbox{Re}{\Sigma}_{XP} \, G_{XP}
   \vphantom{P\!\!\!\!\mbox{\large/}}
                         -\pi^2\,\Gamma_{XP}\,{\cal A}_{XP}\right]
\;,\end{eqnarray}
and
\begin{eqnarray}\label{k8b}\nonumber
\mbox{Tr}\left[\left(
  P\!\!\!\!\mbox{\large/} - M - \cos\Diamond\;\Sigma^\delta_X
  \right)  {\cal A}_{XP}\right]& =&
  \cos\Diamond\; \mbox{Tr}\left[ \vphantom{P\!\!\!\!\mbox{\large/}}
  \mbox{Re}{\Sigma}_{XP}\, {\cal A}_{XP}
                         + \Gamma_{XP} \, G_{XP}\right]\\ 
\mbox{Tr}\left[\left(
  P\!\!\!\!\mbox{\large/} - M - \cos\Diamond\;\Sigma^\delta_X
  \right)  G_{XP}\right]& =&
  \mbox{Tr}\left[1\right] +
  \cos\Diamond\; \mbox{Tr}\left[ \vphantom{P\!\!\!\!\mbox{\large/}}
  \mbox{Re}{\Sigma}_{XP} \, G_{XP}
                         -\pi^2\,\Gamma_{XP}\,{\cal A}_{XP}\right]
\;.\end{eqnarray}
The first pair of these equations is obtained from the
second pair by application of the differential operator $2\tan\Diamond$.
This can be traced back to the fact, that the inverse free
propagator in eqn. \fmref{smix} is a mixture of zeroth and first
order in $\Diamond$, formally
\begin{equation}\label{tand}
\left(P\!\!\!\!\mbox{\large/} -M +
  \frac{{\mathrm i}}{2} \partial\!\!\!\mbox{\large/}_X\right)\,G_{XP}
=\left(1-{\mathrm i}\tan\Diamond\right)\;\left(P\!\!\!\!\mbox{\large/} 
 -M\right)\,G_{XP}
\;.\end{equation}        
When the gradient expansion is truncated without considering this fact,
this mixture may be the source of spurious terms
in transport equations \cite{MH93}.

Obviously, eqn. \fmref{k8b} contains only even powers of the
differential operator $\Diamond$. Solving it in {\em zeroth\/} order
$\Diamond$ therefore gives a spectral function that is correct up to
{\em first\/} order in $\Diamond$. 
\section{Determination of the spectral function}
\begin{figure}
\vspace*{8cm}
\includegraphics{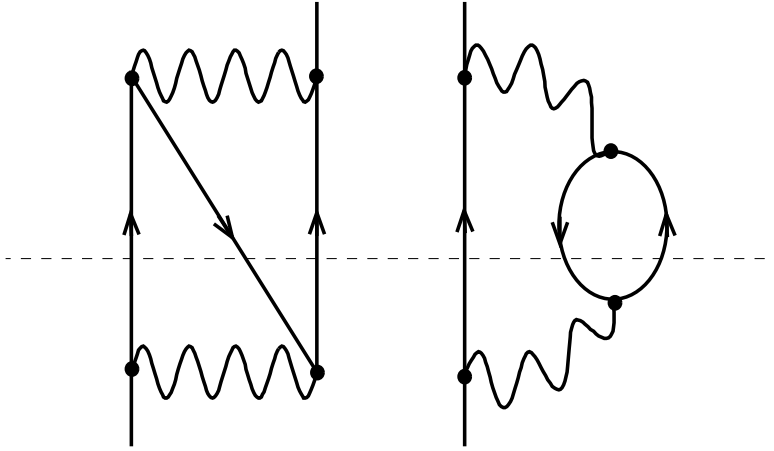}
\caption{Two-loop self energy diagrams corresponding 
         to Born scattering diagrams.}
\label{feynf}
\end{figure}    
We now solve the diagonal components of the Schwinger-Dyson
equation, by making an explicit ansatz for the Dirac spinor
structure of the self energy function:
\begin{equation}\label{svt}
\Sigma^\delta_{X}+\mbox{Re}{\Sigma}_{XP}= S(X,P) + V^\mu(X,P)\gamma_\mu
  +T^{\mu\nu}(X,P)\sigma_{\mu\nu}
\;,\end{equation}
where $S(X,P)$, $V(X,P)$ and $T(X,P)$ stand for (Lorentz) scalar, 
vector and tensor
component of the real part of the self energy,
$\sigma_{\mu\nu} = {\mathrm i}/2\left[\gamma_\mu,\gamma_\nu\right]$
and $T$ is antisymmetric. To avoid confusion with the symbol used for
the fermion propagator, we exploit that the scalar and vector
piece of the real self energy function only appear in 
combination with $M$ and $P$. Hence in the following we make use of
the abbreviations 
\begin{equation}
M^\star     = M + S(X,P)
\;\;\;\;\;\;
P^\star_\nu = P_\nu - V_\nu(X,P) 
\;\end{equation}
for effective mass and momentum. Note, that in contrast to other derivations
of transport equations, the above functions depend on both $X$ and $P$
\cite{MH93}.

Usual derivations of transport equations assume
that the self energy contains
only Lorentz vector and scalar parts \cite{PW88a,BM90}.  It
is however easy to see that for reasonable descriptions of
nuclear dynamics a tensor part of the self energy should be
taken into account. To this end we consider a system, where fermions
are minimally coupled to bosons. In such a model, the
Born diagrams for fermion-fermion scattering are obtained
by cutting the diagrams in figure \ref{feynf} along the
dotted line. If we assume, that the Born terms  are dominant 
contributions to the collision term in ``molecular'' transport equations,
we thus have to take the diagrams of figure \ref{feynf} into account in
our self energy function.

The contribution of the left-hand diagram in the figure to
the fermion self energy then contains products of up to
three $\gamma$-matrices. The threefold product is equivalent to a
pseudovector, which can be excluded in a parity eigenstate. The
twofold product can be converted to a sum of unit matrix and
tensor terms. Thus, if at least two
independent Lorentz vectors can be specified for a momentum
eigenstate, such diagrams contribute tensor parts to the self
energy functions also for homogeneous systems.

Clearly this is the case for a system at finite density, where
the Lorentz invariance is broken: Independent Lorentz vectors are
the (macroscopic) four-current $j^\mu$ of matter
and the momentum of the single-particle state $P_\mu$ (or
equivalently the effective momentum $P^\star_\mu$).
The tensor part of the self energy is then proportional to the commutator
$\left[P^\star_\nu\gamma^\nu,j_\mu\gamma^\mu\right]$.
If one wants to consider phenomena like spin diffusion \cite{MH93}, 
it must not be disregarded.

By virtue of dispersion relations, a similar
decomposition also holds for the imaginary part of the self energy,
\begin{equation}
\Gamma_{XP}= \Gamma_s + \Gamma^\mu_v\gamma_\mu
  +\Gamma^{\mu\nu}_t\sigma_{\mu\nu}
\;,\end{equation}
It follows from the Dyson equations in the previous section,
that the same decomposition then holds
for $G$ and the spectral function:
\begin{eqnarray}\nonumber
G_{XP}&=& G_s + G^\mu_v\gamma_\mu
  +G^{\mu\nu}_t\sigma_{\mu\nu}\\
{\cal A}_{XP}&=& {\cal A}_s + {\cal A}^\mu_v\gamma_\mu
  +{\cal A}^{\mu\nu}_t\sigma_{\mu\nu}
\;.\end{eqnarray}
The ''off-shellness'' of the four momentum $P_\nu$,
and correspondingly the generalized width of the spectral distribution
are given by           
\begin{eqnarray}\label{xqdef}\nonumber
{\cal Y} &=& P^\star_\nu P^{\star\nu}-M^{\star\,2}-2 T_{\mu\nu}T^{\mu\nu}
\\ \nonumber
{\cal X} &=& {\cal Y}+
\pi^2\left(\Gamma_s \Gamma_s - \Gamma_{v,\mu} \Gamma_v^\mu
               + 2 \Gamma_{t,\mu\nu} \Gamma_t^{\mu\nu} \right) 
                 \\
{\cal W} &=& 2\,
  \left(P^\star_\mu\Gamma_v^\mu + M^\star\Gamma_s
  + 2 T_{\mu\nu}\Gamma_t^{\mu\nu} \right)
\;,\end{eqnarray}
and the propagator denominator is
\begin{equation}\label{db2}
{\cal D} = {\cal X}^2 + \pi^2{\cal W}^2
\;.\end{equation}
As was stated after eqn. \fmref{k8b}, we may take the solution
for ${\cal A}$ and $G$ as obtained in homogeneous systems
\cite{PW88a,h91spec}: It is still correct in first order of the
gradient expansion, when the {\em local\/} functions
for effective mass and momentum are inserted.
For reasonably smooth systems, or rather on sufficiently 
large length scales, the equation for the spectral function
therefore is always an algebraic equation.

Explicitly, the solutions are \cite{h91spec}
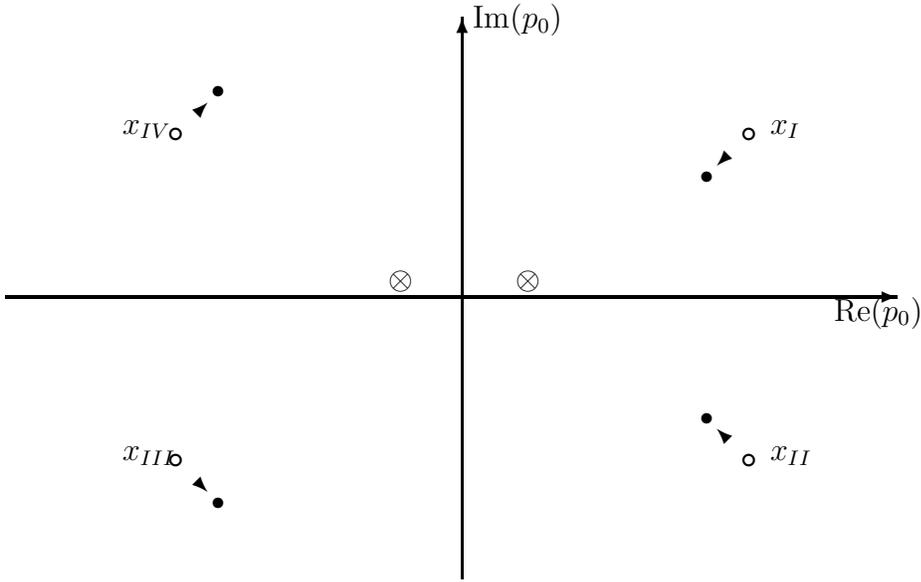
\begin{figure}[t]
\setlength{\unitlength}{0.8bp}
\begin{picture}(420,265)
\thicklines
\put(0,133){\vector(1,0){420}}
\put(215,0){\vector(0,1){265}}
\put(390,122){$\mbox{Re}{(p_0)}$}
\put(220,260){$\mbox{Im}{(p_0)}$}
\put(360,210){$x_I$}
\put(360, 56){$x_{II}$}
\put( 55,210){$x_{IV}$}
\put( 55, 56){$x_{III}$}
\put(350,210){\circle{5}}
\put(350, 56){\circle{5}}
\put( 80,210){\circle{5}}
\put( 80, 56){\circle{5}}
\put(330,190){\circle*{5}}
\put(330, 76){\circle*{5}}
\put(100,230){\circle*{5}}
\put(100, 36){\circle*{5}}
\put(345,205){\vector(-1,-1){10}}
\put(345, 61){\vector(-1, 1){10}}
\put( 85,215){\vector( 1, 1){10}}
\put( 85, 51){\vector( 1,-1){10}}
\put(240,137){$\otimes$}
\put(180,137){$\otimes$}
\end{picture}
\caption{Ghost poles in the perturbative propagator (schematically)}
\label{gho}
$\circ$ are the ghost poles 
of eqn. \fmref{sol5} in a vacuum state, $\bullet$ at nonzero
density. $\otimes$ are the particle poles present in
certain approximations, like e.g. the one discussed in ref. \cite{h92fock}).\\
\hrule
\end{figure}
\begin{eqnarray}\label{sol5n}
G_{XP}&=&\left[
   \left(P\!\!\!\!\mbox{\large/}^\star+M^\star+
T^{\mu\nu}\sigma_{\mu\nu}\right)\,{\cal X}
  -\pi^2
   \left(\Gamma_s-\Gamma^\mu_v\gamma_\mu+
         \Gamma_t^{\mu\nu}\sigma_{\mu\nu}\right)\,{\cal W}\right]
   {\cal D}^{-1}\\ \label{sol5}
{\cal A}_{XP}&=&\left[
   \left(P\!\!\!\!\mbox{\large/}^\star+M^\star+
T^{\mu\nu}\sigma_{\mu\nu}\right)\,{\cal W}
   +
   \left(\Gamma_s-\Gamma^\mu_v\gamma_\mu+
         \Gamma_t^{\mu\nu}\sigma_{\mu\nu}\right)\,{\cal X}\right]
   {\cal D}^{-1}
\;.\end{eqnarray}
These solutions may be obtained in several ways, a straightforward method
is summarized in appendix B. The equations can be combined as
\begin{eqnarray}\label{sol3} \nonumber
&G_{XP}\,\mp\,{\mathrm i}\pi{\cal A}_{XP}&=\;
  \left(P\!\!\!\!\mbox{\large/}-M - \mbox{Re}{\Sigma}_{XP}\,\pm\,{\mathrm i}\pi
  \Gamma_{XP}\right)^{-1}\\
\Leftrightarrow& S^{R,A}_{XP}&=\;
  \left(P\!\!\!\!\mbox{\large/}-M - \Sigma^{R,A}_{XP}\right)^{-1}
\;.\end{eqnarray}
The inverse on the r.h.s. is to be understood as the proper
inverse of a complex 4$\times$4-matrix. 
Hence $G$ and $\pi{\cal A}$ are real and imaginary part of
the same complex function. While this may be used
to determine the functional form of ${\cal A}$ in homogeneous states,
the matrix inversion does not lead to our finding, that ${\cal A}$
in this form is correct also to first order in the gradients.
The solution of the differential equations \fmref{k7a} to this
order therefore is a necessary step of the derivation (see
appendix \ref{spec}).

However, one may not use the above solutions directly. As is discussed
in detail in ref. \cite{h92fock}, a perturbative calculation
of the self energy function $\Sigma$ leads to the invalidation
of the equation for $G$. The reason is, that according
to Weinberg's theorem \cite{W60} the perturbative self energy has
a certain asymptotic behavior, which leads to the appearance
of complex poles in the propagator \fmref{sol3},
see figure \ref{gho}.
        
These Landau ghost poles are unphysical, because they appear
in the wrong complex energy half plane for retarded and advanced
propagator -- and hence have to be removed. In ref. \cite{h92fock},
three strategies were discussed for this removal: Direct subtraction
of the poles, non-perturbative calculation of the self energy function
or the definition of the full propagator as a dispersion integral
\begin{equation}\label{raff}
 G_{XP}\,\mp\,{\mathrm i}\pi{\cal A}_{XP}=
  \int\limits_{-\infty}^\infty\!\!dE\;
   {\cal A}(X;E,\vec{p})\;
   \frac{1}{p_0-E\pm{\mathrm i}\epsilon}
\;,\end{equation}
with $P=(p_0,\vec{p})$. 
It was also shown, that the three strategies
are equivalent, and that furthermore the non-perturbative calculation
of the self energy can be approximated by a form factor giving a better
asymptotic behavior than the one implied by Weinberg's theorem.
For the purpose of the present work, eqn. \fmref{raff} is
considered the {\em definition\/} of $G$, in addition to \fmref{sol5n}.
In coordinate space, the above dispersion relation implies
\begin{equation}\label{raff2}\nonumber
  S^{R,A}_{xy} = \mp 2\pi {\mathrm i}\Theta\left(\pm(x_0-y_0)\right)
        {\cal A}_{xy}
\;.\end{equation}
For simplicity, consider the spectral function
in the case of zero tensor part of the self energy.
Clearly, this spectral function is {\em not generally} of the form
\begin{equation}\label{wro}
{\cal A}_{XP} = \left(P\!\!\!\!\mbox{\large/}^\star + M^\star\right)\,
  \frac{{\cal W}}{\left(P^{\star\,2}-M^{\star\,2}\right)^2+ \pi^2 {\cal W}^2}
\;\end{equation}
with ${\cal W}$ linear in $\Gamma$. This 
equation for the fermionic spectral function is only approximately correct
in case $\Gamma\ll\mbox{Re}{\Sigma}$. Such a functional form
(which was obviously guessed from the non-relativistic result
for the spectral function) has been proposed elsewhere
also for arbitrary self energies \cite{BM90}, 
but it is not reproduced by the complete derivation
carried out here.

The form \fmref{sol3}
of the retarded and advanced propagator implies,
how the limit of zero $\Gamma$ has to be taken, i.e.  as
\begin{equation}\label{noi}
\lim_{\Gamma\rightarrow 0} \Sigma^{R,A}_{XP} =
  \mp i \varepsilon\gamma^0
\;,\end{equation}
to recover the vacuum boundary conditions for the Green functions
\cite{h91spec}.

This can be taken into account by adding an infinitely small part
to the function $\Gamma$. Performing this 
modification in the above equations
then gives for the case of zero imaginary part of the self energy:
\begin{eqnarray}\label{quasi}\nonumber
G_{XP}&=&\frac{
         P\!\!\!\!\mbox{\large/}^\star+M^\star+T^{\mu\nu}\sigma_{\mu\nu}}{{\cal X}}\\
{\cal A}_{XP}&=&
   \left(P\!\!\!\!\mbox{\large/}^\star+M^\star+T^{\mu\nu}\sigma_{\mu\nu}\right)\,
         \mbox{sign}(P^\star_0)\,\delta\left({\cal X}\right)
\;.\end{eqnarray}
In the limit of zero tensor self energy, this turns into
\begin{equation}\label{xpd}
{\cal A}_{XP}= \left( P\!\!\!\!\mbox{\large/}^\star+M^\star\right)
  \,\mbox{sign}(P^\star_0)\;
    \delta\left(P^{\star\,2}-M^{\star\,2}\right)
\;.\end{equation}
Obviously the ``particle''-like pole of the
Green function is renormalized by the derivative
of the self energy.
\section{Transport equation}
As was stated above, the off-diagonal component of the
Schwinger-Dyson equation (in triangular form) is a transport equation.
To study this more closely, we specify the Bogoliubov parameter 
of the transformation \fmref{qptp} as  
\begin{equation}
        n_1=n_2=n_0 = \mbox{const.}
\;\end{equation}
Using eqn. \fmref{dig} and \fmref{dipg}, this
off-diagonal component then has the form
\begin{equation}
S^K=S_0^K + S_0^K\odot\Sigma^A\odot S^A 
          + S_0^R\odot\Sigma^R\odot S^K                                        
          - S_0^R\odot\Sigma^K\odot S^A
\;,\end{equation}
with retarded and advanced components as above and {\em kinetic\/}
components         
\begin{eqnarray}\label{kcd} \nonumber
S_0^K &=& \left( 1- n_0\right)\,S^{12}_0 + n_0\,S^{21}_0\\ \nonumber
S^K   &=& \left( 1- n_0\right)\,S^{12} + n_0\,S^{21}\\   
\Sigma^K &=& \left( 1- n_0\right)\,\Sigma^{12} + 
        n_0\,\Sigma^{21}
\;.\end{eqnarray}
As in the preceding section, 
Dirac differential operators are applied to the off-diagonal
Schwin\-ger-Dyson equation component. They act on the retarded
and advanced propagator as specified in eqn. \fmref{s0sa},
and annihilate the kinetic component $S_0^K$. The kinetic component
of the matrix valued Schwinger-Dyson equation therefore is
\begin{equation}\label{k5} 
\widehat{S}^{-1}_0   S^K_{xy}     =  \Sigma^R_{xz} \odot S^K_{zy}
        -              \Sigma^K_{xz} \odot S^A_{zy}
\;.\end{equation}
Together with its adjoint,
this equation can be expressed entirely in terms of
the off-diagonal components of self energy function
and propagator. These equations then are known as the 
{\em Kadanoff-Baym equations\/}, see ref. \cite{KB62}.

The next step is to transform to the mixed (Wigner) representation,
and to insert the gradient expansion,
\begin{eqnarray}\label{k8} \nonumber
&&\left(P\!\!\!\!\mbox{\large/} -M +\frac{{\mathrm i}}{2}
  \partial\!\!\!\mbox{\large/}_X - \exp(-{\mathrm i}\Diamond)\,\Sigma^\delta_X
  \right)  S^K_{XP}   \\
&&\;\;\;\;\;\;=
\exp(-{\mathrm i}\Diamond)  \left[ 
       \left(\mbox{Re}{\Sigma}_{XP}-{\mathrm i}\pi\Gamma_{XP}\right)
         S^K_{XP}  + \Sigma^K_{XP}\left(G_{XP}+{\mathrm i}\pi {\cal A}_{XP}
                    \right)\right]
\end{eqnarray}
The separation of real and imaginary part is a trivial step.
However, for comparison with existing derivations we perform it
explicitly and obtain the two equations
\begin{eqnarray}\label{tpe1a} \nonumber
\mbox{Tr}\left[\left(
  \partial\!\!\!\mbox{\large/}_X + 2 \sin\Diamond\;\left(\Sigma^\delta_X
  +\mbox{Re}{\Sigma}_{XP}\right)
   + \cos\Diamond\;2\pi\Gamma_{XP}
  \right)  S^K_{XP}\right]& = \\
   2{\mathrm i} \mbox{Tr}\left[
             {\mathrm i}\sin\Diamond\;\Sigma^K_{XP} \, G_{XP}
             - \cos\Diamond\;\Sigma^K_{XP} \, 
    {\mathrm i}\pi{\cal A}_{XP}\right]&
\end{eqnarray}
\begin{eqnarray}\label{tpe1b}\nonumber                        
\mbox{Tr}\left[\left(
  P\!\!\!\!\mbox{\large/} - M - \cos\Diamond\;\left(\Sigma^\delta_X
  +\mbox{Re}{\Sigma}_{XP}\right)
   + \sin\Diamond\;\pi\Gamma_{XP}
  \right)  S^K_{XP}\right]& = \\
        \mbox{Tr}\left[
               \cos\Diamond\;\Sigma^K_{XP} \, G_{XP}
            -{\mathrm i}\sin\Diamond\;\Sigma^K_{XP} \, 
  {\mathrm i}\pi{\cal A}_{XP}\right]&
\;.\end{eqnarray}
It is obvious, that these equations are not independent:
As was already observed for the retarded and advanced equation,
\fmref{tpe1a} follows from \fmref{tpe1b} by the application of a differential
operator. Usually only eqn. \fmref{tpe1a} is processed
further, leading to certain disadvantages that we will avoid below.

The final result of the processing will be an
equation for the quantity $N_{XP}$ defined by
\begin{equation}\label{nde}
\left(1-N_{XP}\right)\,S_{XP}^{12} + N_{XP}\,S_{XP}^{21} =0
\,.\end{equation}  
For the purpose of the present paper $N_{XP}$ is taken as a scalar function,
it has the physical meaning of a generalized phase-space distribution function
\cite{L90}. The description of Phenomena 
like spin diffusion require to use a Dirac matrix valued $N_{XP}$.

It follows according to eqns. \fmref{raff2} and \fmref{kcd}, that
\begin{equation}\label{gkk}
S_{XP}^K = 2\pi{\mathrm i}\,\left(N_{XP} - n_0\right)\,{\cal A}_{XP}
\;.\end{equation}                                  
Equation \fmref{nde} has two aspects which are important for the
understanding of transport phenomena. First, consider it in
the equilibrium case: No dependence on $X$ is present. The equation
is then equivalent to the Kubo-Martin-Schwinger boundary condition
for the propagator \cite{KMS} -- which implies the proper 
asymptotic property for the quantity $N_{XP}$ we are searching.

The second aspect of the above equation only makes sense in the
context of thermo field dynamics (TFD). There, it is the
condition for a {\em diagonalization} of the full propagator
by (space-time dependent) Bogoliubov transformations, i.e.,
according to eqn. \fmref{dig}
\begin{equation}
{\cal B}(N_{XP})\,\tau_3\,S_{XP}\,({\cal B}(N_{XP}))^{-1}=
\left({\array{rr} G_{XP}-{\mathrm i}\pi{\cal A}_{XP} & \\
 & G_{XP}+{\mathrm i}\pi{\cal A}_{XP} \endarray}\right)
\;.\end{equation}
Hence, the full propagator (in mixed representation) is
diagonalized by a space-time dependent Bogoliubov transformation.
One may insert this propagator into the 
Schwin\-ger-Dy\-son equation --  the action of the Dirac
differential operator on the function $N_{XP}$ would have to be considered
then. According to eqn. \fmref{tnf}, this would only affect
the off-diagonal component. The strategy of diagonalizing the 
full non-equilibrium propagator is therefore {\em equivalent\/} 
to the processing of the Schwinger-Dyson equation as done here. 

For the following, we furthermore
define a ``pseudo-equilibrium'' distribution
function: The property \fmref{sse} of the
self energy function allows to diagonalize its 2$\times$2 matrix structure
by a Bogoliubov transformation (see eqn. \fmref{dipi}) with
a parameter $N^0_{XP}$ such that
\begin{equation}\label{psef}
\Sigma^{12}_{XP} =  2\pi{\mathrm i} N^0_{XP}\,\Gamma_{XP}\;\;\;\;\;\;\;
\Sigma^{21}_{XP} =  2\pi{\mathrm i} \left(N^0_{XP}-1\right)\,\Gamma_{XP}
\;.\end{equation}
It is easy to show, that in {\em equilibrium\/} states
$N^0$ is identical to $N$ as defined in \fmref{nde} \cite{hu92}. For
non-equilibrium states however, $N^0$ may vary with time and be quite different
from $N_{XP}$. Similar to eqn. \fmref{gkk} we obtain
\begin{equation}\label{gkk2}
\Sigma_{XP}^K = 2\pi{\mathrm i}\,\left(N^0_{XP} - n_0\right)\,\Gamma_{XP}
\;.\end{equation}                                  
We now return to the transport equation \fmref{k8},
perform a gradient expansion to first order in $\Diamond$ and insert 
the above definitions:
\begin{eqnarray}\label{t1}\nonumber                        
\mbox{Tr}\left[
  \left(1-{\mathrm i}\Diamond\right)\,
  \left\{\vphantom{\int} P\!\!\!\!\mbox{\large/} - M - 
   (\Sigma^\delta_X+\mbox{Re}{\Sigma}_{XP})+{\mathrm i}\pi\Gamma_{XP}\right\}\,
  \left\{\vphantom{\int}\left(N_{XP}-n_0\right){\cal A}_{XP}\right\}
\right]& = \\
\mbox{Tr}\left[
  \left(1-{\mathrm i}\Diamond\right)\,
  \left\{\vphantom{\int}\left(N^0_{XP}-n_0\right)\Gamma_{XP}\right\}\,
  \left\{\vphantom{\int}G_{XP}+{\mathrm i}\pi{\cal A}_{XP}\right\}
\right]&
\;.\end{eqnarray}
The curly brackets contain the quantities acted upon by the
$\Diamond$ operator according to \fmref{dia1}.

The derivative of ${\cal A}$ on the left side is eliminated using
\fmref{k8a}, and the non-derivative terms on the left side
are switched over to the right side using \fmref{k8b}. All terms
involving $n_0$ drop out correctly, and the resulting
full quantum transport equation is
\begin{eqnarray}\label{t2}\nonumber                        
\mbox{Tr}\left[ {\cal A}_{XP}\,
  \left(-{\mathrm i}\Diamond\right)\,
  \left\{\vphantom{\int} P\!\!\!\!\mbox{\large/} - M - 
   (\Sigma^\delta_X+\mbox{Re}{\Sigma}_{XP})+{\mathrm i}\pi\Gamma_{XP}\right\}\,
  \left\{\vphantom{\int}N_{XP}\right\}
\right]& = \\ \nonumber
\mbox{Tr}\left[
  -N_{XP}\,\left(1-{\mathrm i}\Diamond\right)\,
  \left\{\vphantom{\int}\Gamma_{XP}\right\}\,
  \left\{\vphantom{\int}G_{XP}+{\mathrm i}\pi{\cal A}_{XP}\right\}
\right.\\
\left.\hphantom{\mbox{Tr}\left[\right.N_{XP}\,}
  +\left(1-{\mathrm i}\Diamond\right)\,
  \left\{\vphantom{\int}N^0_{XP}\,\Gamma_{XP}\right\}\,
  \left\{\vphantom{\int}G_{XP}+{\mathrm i}\pi{\cal A}_{XP}\right\}
\right]&
\;.\end{eqnarray}
\section{Vlasov equation}
As the next step, we rederive the ``standard'' Vlasov equation without 
collision term. To this end, the right side of eqn. \fmref{t2}
is dropped completely. Furthermore, the traditional 
quasi-particle approximation is made: The spectral function is chosen
according to \fmref{xpd}. Note, that according to eqn. \fmref{raff}
this is equivalent to an energy-independent real self energy
function in Wigner representation. In coordinate space, such
a picture is completely consistent only in an approximation,
where all self energy functions are local in space and time.
In a relativistic model this is the case only for the Hartree approximation,
nonrelativistically one may also add the exchange (Fock) self energy
\cite{D84a}.

In leaving the covariant notation, we set $P=(p_0,\vec{p})$
and $X=(t,\vec{x})$. Furthermore we assume,
that the spectral $\delta$-function \fmref{xpd} 
has at least one pole along
the real $p_0$-axis, i.e. that one can rewrite it as
\begin{equation}
\delta\left(P^{\star\,2}-M^{\star\,2}\right) = \sum\limits_i
         \delta\left(p_0-E_i(\vec{x},\vec{P},t)\right)
   \;\left|\frac{\partial (P^{\star\,2}-M^{\star\,2})}{\partial p_0}\right|^{-1}
\;,\end{equation}
where $E_i(\vec{x},\vec{P},t)$ is the (space-time and momentum
dependent) generalized energy of the corresponding ``particle''-like
state.

Abbreviating the function $N_{XP}$ at
each of the poles as $N_i(\vec{x},\vec{p},t)$,
an integration over the energy variable then eliminates
the $\delta$-function as well as the energy derivative, and one
obtains from eqn. \fmref{t2}
\begin{eqnarray}\label{VL2}\nonumber
& \sum\limits_i&\left( \frac{\partial N_i(\vec{x},\vec{p},t)}{\partial t}
\right. \\
 && \left.+\frac{\partial E_i(\vec{x},\vec{p},t)}{\partial \vec{p}}
   \frac{\partial N_i(\vec{x},\vec{p},t)}{\partial \vec{x}}
  -\frac{\partial E_i(\vec{x},\vec{p},t)}{\partial \vec{x}}
   \frac{\partial N_i(\vec{x},\vec{p},t)}{\partial \vec{p}}\right) = 0
\;.\end{eqnarray}
This is the Vlasov (or mean-field)  quantum
transport equation in standard notation: The system evolves in time  
free of dissipation. 

It is quite instructive to rewrite this in a covariant notation.
Using the spectral function defined in eqn.
\fmref{xpd} and ${\cal Y}$ as defined in \fmref{xqdef}, 
one can transform equation \fmref{t2} in a few lines
into
\begin{equation}\label{VL1}
 \delta({\cal Y})
 \,\left\{ \frac{\partial {\cal Y}}{\partial P_\mu}\,
        \frac{\partial N_{XP}  }{\partial X^\mu}  \;
     -\;\frac{\partial {\cal Y}}{\partial X_\mu}\,
        \frac{\partial N_{XP}  }{\partial P^\mu}  \right\}=0
\;.\end{equation}
This may be integrated over the variable ${\cal Y}$,
hence the covariant Vlasov equation amounts to the vanishing of the expression
in curly brackets on the hypersurface ${\cal Y}=0$
\cite[pp.438]{L90}. The characteristic curves of
this partial differential equation are parameterized by $s$,
\begin{equation}\label{ce}
\frac{d X^\mu}{d s} = \frac{\partial {\cal Y}}{\partial P_\mu}
\;\;\;\;\;\;\;\;\;
\frac{d P^\mu}{d s} = -\frac{\partial {\cal Y}}{\partial X_\mu}
\;.\end{equation}
Hence eqn. \fmref{VL1} states, that the total derivative 
of $N_{XP}$ with respect to $s$ vanishes on the hypersurface 
defined by ${\cal Y}=0$. The characteristic
equations, together with the constraint, are Hamilton's
equations for an infinitesimal phase-space ``fluid element'' with proper time
proportional to $s$. In numerical simulations, these two equations
determine the test-particle trajectories \cite {GCM93}.

Note, that the sum over the poles appearing in
\fmref{VL2} is also inherent to \fmref{VL1}: To make the latter
meaningful, one has to solve the constraint equation
${\cal Y}=0$, and in general will find that the hypersurface it
defines consists of two or more disjoint parts.
\section{Transport equation with quantum effects}
The second step in the study of the quantum transport equation
\fmref{t2} consists of neglecting only the {\em gradient\/}
terms on the right hand side. Real and imaginary part separate easily,
\begin{eqnarray}\label{c1}\nonumber                        
\mbox{Tr}\left[ {\cal A}_{XP}\,
  \left(2\Diamond\right)\,
  \left\{\vphantom{\int} \pi\Gamma_{XP})\right\}\,
  \left\{\vphantom{\int}N_{XP}\right\}
\right]&= \\  
  -2 \,\left(N_{XP}-N^0_{XP}\right)&\!\!
\mbox{Tr}\left[\Gamma_{XP}\,\vphantom{\int}G_{XP}\right] \\ \nonumber
\label{t3}
\mbox{Tr}\left[ {\cal A}_{XP}\,
  \left(2\Diamond\right)\,
  \left\{\vphantom{\int} P\!\!\!\!\mbox{\large/} - M - 
   (\Sigma^\delta_X+\mbox{Re}{\Sigma}_{XP})\right\}\,
  \left\{\vphantom{\int}N_{XP}\right\}
\right]& =\\
 2\pi\,\left(N_{XP}-N^0_{XP}\right)&\!\!
\mbox{Tr}\left[
  \Gamma_{XP}\,\vphantom{\int}{\cal A}_{XP}
\right]
\;.\end{eqnarray}
Obviously, this is a transport
equation in {\em relaxation time approximation\/}: The derivative of
$N_{XP}$ is proportional to the difference of $N_{XP}$ and $N^0_{XP}$. 

Up to now the spectral function is completely general.
However, to study the consequences
of the off-shell effects in this equation, we have to relate the
it as closely as possible to the 
Vlasov equation. This achieved by using the {\em approximate\/}  
spectral function \fmref{wro}, which amounts to
an expansion of the complete solution \fmref{sol5} up to
lowest order in $\Gamma$ and neglection of the tensor parts.
It is inserted in eqn. \fmref{t3}, i.e., into the imaginary
part of the full transport equation.
One obtains, after calculation of the trace over the Dirac matrices
\begin{equation}\label{tpe6}\nonumber
\left(\frac{{\cal W}}{{\cal Y}^2+\pi^2{\cal W}^2}\right)\,
 \left\{ \frac{\partial {\cal Y}}{\partial P_\mu}\,
        \frac{\partial N_{XP}}{\partial X^\mu} -
        \frac{\partial {\cal Y}}{\partial X_\mu}\,
        \frac{\partial N_{XP}}{\partial P^\mu} 
  \;+\;2\pi\,{\cal W}  \left(N_{XP}- N^0_{XP}\right)\right\}=0
\;.\end{equation}
While one may consider this equation for each value of $p_0$,
it obtains a more specific meaning when we perform
the integration over ${\cal Y}$. The insertion of the spectral function
\fmref{wro} already implies a sufficiently small ${\cal W}$,
hence the overall factor outside the curly brackets can be
approximated by a $\delta$-function in this case.

Then the physical interpretation of ${\cal W}$ is that of a spectral
width multiplied by twice the energy of an almost particle-like
mode, and the energy of this mode is the positive solution
of ${\cal Y}=0$ (see \fmref{xqdef}). For this mode we obtain 
\begin{equation}\label{VL3}
\left(\frac{d N_{XP}}{d s}\right)_{{\cal Y}=0} =
 \left( \frac{\partial {\cal Y}}{\partial P_\mu}\,
        \frac{\partial N_{XP}  }{\partial X^\mu}  \;
     -\;\frac{\partial {\cal Y}}{\partial X_\mu}\,
        \frac{\partial N_{XP}  }{\partial P^\mu}  \right)_{
         {\cal Y}=0} = -2\pi\, {\cal W}
        \left(N_{XP}-N^0_{XP}\right)_{
         {\cal Y}=0}
\end{equation}
As argued above $s$ is proportional to the proper time along
the characteristic curves of the Vlasov equation, and these characteristic
curves are the hamiltonian trajectories of an infinitesimal 
phase-space cell, see \fmref{ce}. 

The standard Vlasov equation states, 
that the occupation probability of this phase space cell does not change with 
proper time.  Conversely the equation \fmref{VL3} expresses, that
such a change of the distribution function occurs in case
the spectral function peak has a nonzero width. 
Hence, instead of the Vlasov equation, the result is a kinetic
equation with dissipation.

An equation similar to \fmref{VL3} has been obtained by several authors
in spatially homogeneous fermionic and bosonic systems
\cite{NBH93,WR93,hu92,BM90,EW92,E92,habil}. For the
purpose of understanding relaxation phenomena it is of tremendous
importance: It implies, that without an imaginary part of
the self energy function and correspondingly a nonzero spectral width
there is no relaxation in a quantum system. 

The interpretation of trajectories in phase space is more complicated
when proceeding to the full spectral function \fmref{sol5}. 
Also, when using only eqn. \fmref{t3} one obtains 
ugly expressions of the type $\Gamma_s\partial_XM^\star\dots$
Hence one has to combine it with the real part of eqn. \fmref{t2},
which is written explicitly in \fmref{c1}. The latter
then contains terms of the type $M^\star\partial_X\Gamma_s\dots$

The result of this combination is written down only for the
sake of completeness: It demonstrates how the 
{\em quantum\/} transport equation including dissipation may be written
in covariant form. Using the abbreviations defined in \fmref{xqdef},
one obtains
\begin{eqnarray}\label{c2}\nonumber
&&-\frac{{\cal X}}{\pi {\cal W}}\,\times\mbox{eqn. \fmref{c1}} + 
\mbox{eqn. \fmref{t3}}\Rightarrow 
\;\left(\frac{1}{{\cal X}^2+\pi^2{\cal W}^2}\right)\; 
\times\\ \nonumber
&& \left\{ {\cal W}\, \left(\frac{\partial {\cal Y}}{\partial P_\mu}\,
        \frac{\partial N_{XP}}{\partial X^\mu} -
        \frac{\partial {\cal Y}}{\partial X_\mu}\,
        \frac{\partial N_{XP}}{\partial P^\mu}\right)\right. 
-{\cal X}\,\left(\frac{\partial {\cal W}}{\partial P_\mu}\,
        \frac{\partial N_{XP}}{\partial X^\mu} -
        \frac{\partial {\cal W}}{\partial X_\mu}\,
        \frac{\partial N_{XP}}{\partial P^\mu}\right) \\ \nonumber
&&-\frac{{\cal X}^2}{\pi^2{\cal W}}\,\left(
        \frac{\partial {(\cal X-\cal Y)}}{\partial P_\mu}\,
        \frac{\partial N_{XP}}{\partial X^\mu} -
        \frac{\partial {(\cal X-\cal Y)}}{\partial X_\mu}\,
        \frac{\partial N_{XP}}{\partial P^\mu}\right) \\
&&+\left.\vphantom{\int}
       2\pi\,\left(N_{XP}- N^0_{XP}\right)
\,\left(\left({\cal W}^2+{\cal X}^2/\pi^2\right)
  + 2\left({\cal X} - \frac{{\cal X}^2}{{\cal W}}\right)
     \left({\cal X-\cal Y}\right)\right)
\right\}=0
\;.\end{eqnarray}
As before, this equation can be studied around a pronounced
peak in the spectral function. To first order in the peak width,
eqn. \fmref{tpe6} is recovered.

Finally we put together all knowledge assembled so far, i.e.,
the quantum transport equation \fmref{t2} is used without
further approximation. The crucial problem is now to determine
the role of the gradient terms neglected in the 
derivation of \fmref{c2}.

The right side of \fmref{t2}, which contains these terms,
has the general structure
\begin{equation}\label{t4} 
{\cal C}_{XP}=\left(1- {\mathrm i}\Diamond\right)\;
\widetilde{\Gamma}_{XP} \left( G_{XP}
               +   {\mathrm i}\pi{\cal A}_{XP}\right) 
\approx\exp(-{\mathrm i}\Diamond)\;
\widetilde{\Gamma}_{XP} \left( G_{XP}
               +   {\mathrm i}\pi{\cal A}_{XP}\right) 
\;.\end{equation}
$\widetilde{\Gamma}$ has an obvious meaning to be read 
off from \fmref{t2}, resembling either $\Gamma_{XP}$ or
$N^0_{XP}\Gamma_{XP}$.

In appendix \ref{dik} of the present paper it is shown,
that if and only if the two terms on the right side are related
through the dispersion relation \fmref{raff}, one may express
${\cal C}_{XP}$ up to first order in the gradient expansion as
\begin{eqnarray} \label{t5} \nonumber
{\cal C}_{XP}
&=&\widetilde{\Gamma}_{XP} \left( G_{XP}
               +   {\mathrm i}\pi{\cal A}_{XP}\right) \\
&+&
\int\!\!d\tau\,dE\;
\Theta(-\tau)\,\exp(-{\mathrm i}\tau E)\;
   \Diamond\,\widetilde{\Gamma}(t+\tau/2,\vec{x};P)
   \;{\cal A}(X;p_0+E,\vec{p})
\;\end{eqnarray}
with $X=(t,\vec{x})$ and $P=(p_0,\vec{p})$.
Hence the {\em dissipative kernel\/} of eqn. \fmref{t4} can be split into 
the time-local piece we have considered already in eqn. \fmref{t3},
and an integral over the {\em past\/} of the system.
While the original Schwinger-Dyson equation required a time integration 
reaching into the future of the system, the present equation is strictly
causal.

Using the explicit meaning of $\widetilde{\Gamma}$ from
eqn. \fmref{t2}, the final result of the
present work is the equation
\begin{eqnarray}\label{tpefin}\nonumber                        
&&\mbox{Tr}\left[ {\cal A}_{XP}\,
  \left(2\Diamond\right)\,
  \left\{\vphantom{\int} P\!\!\!\!\mbox{\large/} - M - 
   (\Sigma^\delta_X+\mbox{Re}{\Sigma}_{XP})\right\}\,
  \left\{\vphantom{\int}N_{XP}\right\}\right]  \\ \nonumber 
&&\;\;\;\;=\;2\pi\,\left(N_{XP}-N^0_{XP}\right)\,
\mbox{Tr}\left[\Gamma_{XP}\,\vphantom{\int}{\cal A}_{XP}\right]\\ \nonumber
&&\;\;\;\;+\;N_{XP}\; \int\!\!d\tau\,dE\;
\Theta(-\tau)\,\sin(\tau E)\,\mbox{Tr}\left[
 \vphantom{\int}\left(2\Diamond\right)\,
\Gamma(t+\tau/2,\vec{x};P)\;{\cal A}(X;p_0+E,\vec{p})\right]\\
&&\;\;\;\;-\;\int\!\!d\tau\,dE\;
\Theta(-\tau)\,\sin(\tau E)\,\mbox{Tr}\left[
 \vphantom{\int}\left(2\Diamond\right)\,
  (N^0\,\Gamma)(t+\tau/2,\vec{x};P)\;
  {\cal A}(X;p_0+E,\vec{p})\right]
\;.\end{eqnarray} 
The physical interpretation of the integral over the past history of
the system is obvious: It leads to the memory effects observed 
in transport theory, and relates these to the spectral width
of the ''particles'' propagating in the medium.

Note, that one may insert a diagrammatic
expansion for the self energy function, and thereby for $\Gamma_{XP}$
and $N^0_{XP}\Gamma_{XP}$. Consider e.g. the one depicted in figure 1:
In quasi-particle approximation it leads to a standard
collision integral on the right side of the above equation. 
However, beyond the quasi-particle approximation the oscillating
factor $\sin(\tau E)$ may lead to enhancement as well as
suppression of relaxation. 
\section{Conclusions} 
In the present work,
an equation of motion for the relativistic one-particle phase-space 
distribution function of a fermion system was derived within a real time 
Green function formalism. The only approximation made is an expansion to 
first order in the gradients present in a non-equilibrium system.
In particular, no quasi-particle approximation was introduced:
The full spectral information to the same  order in the gradient
expansion is retained in the dissipative equation \fmref{tpefin}.
This places the present derivation 
in the gap between quantum field theory and traditional transport equations.

A result of this study is the observation, that dissipation
occurs if and only if the spectral function deviates from
a $\delta$-function, i.e., if the imaginary part of the
self energy function is nonzero. For strongly peaked
spectral functions, which one might associate with ``particles'',
this implies that dissipation is equivalent to a 
finite lifetime of the excitation.

Such an interpretation is fully consistent with standard
``molecular'' transport theory: The 
Uehling-Uhlenbeck collision term in ``molecular''
transport equations is diagrammatically given by Born diagrams
\cite{GCM93}.
Their closure to a self energy function, as depicted in figure 1,
gives the lowest order perturbative contribution to the width
of the quasi-particle pole in the propagator . 
This result makes the quasi-particle approach to 
the simulation of nuclear collisions quite dubious.

Close to equilibrium, the dissipative driving force is proportional
to the difference between $N_{XP}$ and $N^0_{XP}$, i.e., between the
time local distribution function diagonalizing the propagator and 
the one diagonalizing the self energy function. 
The strategy of diagonalizing the 
full non-equilibrium propagator by thermal Bogoliubov transformations
is therefore {\em equivalent\/} to the solution of the transport equation.
One may put this into the form of a statement 
about the TFD formalism: The basic idea of thermo field dynamics, 
the Bogoliubov transformation,
achieves the proper separation of spectral and statistical information 
even in a non-equilibrium system \cite{hu92,habil}. 

To first order in the gradients the spectral function is the
solution of an algebraic equation, eqn. \fmref{sol5}.
This may be cast together with the definition
of the real part of the full propagator by dispersion integral,
eqn. \fmref{raff} and an ansatz to determine 
the self energy from the propagator. Such an ansatz was not discussed
in the present work, it necessitates a perturbative expansion
in terms of the full propagators that has been formulated
elsewhere \cite{EW92,E92,habil}.
We then have at hand a closed system of equations for the 
Green function of the interacting fermion system, with controlled
approximations. 

The rewriting of the dissipative kernel in terms of the spectral function
and the self energy was achieved by using a dispersion integral.
This necessitates a proper understanding of the analytical structure
of the retarded and advanced Green function of the system, {\em before}
following the time evolution of occupation probabilities.
Hence, the proper ghost-free definition
of the real part of the propagator, as outlined before and
discussed in ref. \cite{h92fock}, is an essential
ingredient of this step. 

The previously made artificial distinction between the spectral width 
due to collisions and a ``natural'' spectral width, 
like e.g. for resonances \cite{M86}, does not arise in the present work. 
However, a distinction is possible between
time-local relaxation and memory effects present in the system.
The influence exerted by the memory kernel of eqn. \fmref{tpefin}
depends on the details of spectral function and time evolution:
One may not estimate its effect on the net relaxation time {\em a priori}.

In ref. \cite{WR93} it was emphasized, that even in a simple model for 
relativistic heavy-ion collisions a slowdown as well as an enhancement
may occur through memory effects. In \cite{D84a} an average 
slow-down was found in a similar non-relativistic framework,
by comparing the solution of the Schwinger-Dyson equation with
a ``molecular'' transport equation. 
Accordingly, the next logical step will be a numerical study
of the derived transport equation \fmref{tpefin}, but this is
beyond the scope of the present work.
\subsection*{Acknowledgement}
I wish to thank R.Malfliet and S.Mrowczynski for their valuable
comments on the first version of the present work, which was
{\em not\/} published in 1992. I am also very much indebted to
Ch.van Weert for constructive criticism.
\clearpage 
\appendix
\section{Thermal Bogoliubov matrix}\label{ling}
The thermal Bogoliubov transformations of thermo field dynamics exist in
various parameterizations. As has been pointed out in ref. \cite{hu92,Ubook},
the special choice
$\alpha=1$ and $s=1/2\log(1+\sigma n)$ makes them
linear in a single parameter $n$, for fermionic fields
the explicit form is given in eqn. \fmref{lc}. It follows e.g., that
\begin{equation}\label{tnf}
({\cal B}(n))^{-1}\,\frac{d}{d t}{\cal B}(n) = 
\;({\cal B}(n))^{-1}\,
\left( { \array{lr}
         0& -1 \\
         0&  0  \endarray} \right)\,\left(\frac{d n}{d t}\right)
\,{\cal B}(n)
\;.\end{equation}
Obviously, a time-dependence of the Bogoliubov parameter $n$
only affects the off-diagonal part of the Schwinger-Dyson equation.

Now let $\Sigma$ be any $2\times 2$ matrix which fulfills the linear 
relation \fmref{sse}. Then one obtains
\begin{equation}\label{dipi} 
{\cal B}(n_1)\,
\Sigma\tau_3\,({\cal B}(n_2))^{-1}=
\left({\array{rr} \Sigma^R &\;
  \Sigma^{11}(n_2-n_1)-(1 -n_2)\Sigma^{12}-n_1\Sigma^{21}\\
                          & \Sigma^A \endarray}\right)
\;.\end{equation}
Quite similar relations exist for propagator matrices:
Let $S$ be any $2\times 2$ matrix which fulfills the linear relation
\fmref{sme}. Then 
\begin{equation}\label{dig}
{\cal B}(n_1)\,\tau_3\,S\,({\cal B}(n_2))^{-1}=
\left({\array{rr} S^R &\;
  S^{11}(n_2-n_1)+(1-n_2)S^{12}+n_1 S^{21}\\
                          & S^A \endarray}\right)
\;.\end{equation}
For the purpose of deriving transport equations it is furthermore
advantageous to exploit a similar relation for the product of
three such matrices. Let $S_0$ also be a propagator matrix,
whose matrix elements satisfy the relations \fmref{sme}.
Then one obtains
\begin{equation}\label{dipg} 
{\cal B}(n_1)\,\tau_3\,S_0\,\Sigma\,S\,({\cal B}(n_2))^{-1}=\\
\left({\array{rr} S^R_0\,\Sigma^R\,S^R & X \\
             & S^A_0\,\Sigma^A\,S^A \endarray}\right)
\;,\end{equation}
where we have used the definition of retarded and advanced components
as found in eqn. \fmref{sra}, and
\begin{eqnarray} \nonumber
X &=&
  S^K_0\,\Sigma^A\,S^A +
  S^R_0\,\Sigma^R\,S^K \\
&+&
  S^R_0\,\left((n_2-n_1) \Sigma^{11}
                    +(n_2-1) \Sigma^{12} -n_1 \Sigma^{21}
  \right)\,S^A
\;\end{eqnarray}
with the kinetic components defined in eqn. \fmref{kcd}.
Corresponding relations hold for bosonic Bogoliubov transformations,
see ref. \cite{hu92}.
\section{Fermionic spectral function}\label{spec}
In this section, the equations \fmref{k7a} are solved using
\fmref{svt} for the spinor structure of
the self energy function.
First, the trace is taken over the 
Dirac matrices:        
\begin{eqnarray}\label{ss1}\nonumber
\cos\Diamond\;\left(P_\mu^\star G_v^\mu - M^\star G_s-
2 T_{\mu\nu}G_t^{\mu\nu}\right)&=&
1 - \pi^2
  \cos\Diamond\;\left( \Gamma_s\,{\cal A}_s+\Gamma_{v,\mu}\,{\cal A}_v^\mu
  +2 \Gamma_{t,\mu\nu}{\cal A}_t^{\mu\nu}
  \right)\\ 
\cos\Diamond\;\left(P_\mu^\star {\cal A}_v^\mu -
  M^\star {\cal A}_s-2 T_{\mu\nu}{\cal A}_t^{\mu\nu}\right)&=&
  \cos\Diamond\;\left( \Gamma_s\,G_s+\Gamma_{v,\mu}\,G_v^\mu
  +2 \Gamma_{t,\mu\nu}G_t^{\mu\nu}
  \right)
\;.\end{eqnarray}
Multiplication with
$\gamma^\nu$ before taking the trace gives the next equations.
To eliminate the term $\partial_\mu\,G_t^{\mu\nu}$ one uses \fmref{tand},
\begin{eqnarray}\label{sv1} \nonumber
&&\cos\Diamond\;\left(P^{\nu \star}G_s-M^\star G_v^\nu\right) 
+2\sin\Diamond\;\left( P_\mu^\star G_t^{\mu\nu} - T^{\nu\mu}\,G_{v,\mu}\right)
\\ \nonumber
&&\;\;\;\;\;\;  = - \pi^2
  \cos\Diamond \;\left( 
   \Gamma_s\,{\cal A}_v^\nu+\Gamma_v^\nu\,{\cal A}_s \right)
  - 2\pi^2
  \sin\Diamond\;\left( \Gamma_t^{\nu\mu}\,{\cal A}_{v,\mu}+
                   \Gamma_{v,\mu}\,{\cal A}_t^{\mu\nu} \right) \\ \nonumber
&&\cos\Diamond\;\left(P^{\nu \star}{\cal A}_s-M^\star {\cal A}_v^\nu
\right)+2\sin\Diamond\;\left( P_\mu^\star {\cal A}_t^{\mu\nu}-
  T^{\nu\mu}\,{\cal A}_{v,\mu}\right) \\
&&\;\;\;\;\;\;  =
  \cos\Diamond \;\left( \Gamma_s\,G_v^\nu+\Gamma_v^\nu\,G_s \right)
  + 2\sin\Diamond\; \left( \Gamma_t^{\nu\mu}\,G_{v,\mu}+
                   \Gamma_{v,\mu}\,G_t^{\mu\nu} \right)
\;.\end{eqnarray}
The remaining spin components are obtained by multiplying with
${\mathrm i}\gamma^5\gamma^\alpha$ before taking the trace:
\begin{eqnarray}\label{st1} \nonumber
\epsilon^{\alpha\nu\sigma\tau}\,\cos\Diamond\;\left(
  P_\nu^\star G_{t,\sigma\tau}-T_{\sigma\tau}\,G_{ ,\nu}\right)&=&
  -\pi^2
  \epsilon^{\alpha\nu\sigma\tau}
  \cos\Diamond\;\left( \Gamma_{t,\sigma\tau}\,{\cal A}_{v,\nu}+
                   \Gamma_{v,\nu}\,{\cal A}_{t,\sigma\tau} \right) \\
\nonumber
\epsilon^{\alpha\nu\sigma\tau}\,\cos\Diamond\;\left(
  P_\nu^\star{\cal A}_{t,\sigma\tau}-T_{\sigma\tau}\,{\cal A}_{v,\nu}\right)
&=&
  \epsilon^{\alpha\nu\sigma\tau}\,
  \cos\Diamond\;\left( \Gamma_{t,\sigma\tau}\,G_{v,\nu}+
                   \Gamma_{v,\nu}\,G_{t,\sigma\tau} \right)
\;,\end{eqnarray}
where
$\epsilon^{\alpha\nu\sigma\tau}$ is the total antisymmetric tensor
in four dimensions. Since $X$ and $P$ are independent variables,
the $\sin\Diamond$-terms from equation \fmref{sv1} cancel.
Furthermore, in first order gradient expansion $\cos\Diamond=1$.
Reduced spinorial components are introduced as
\begin{equation}
a_s\;=\; \frac{{\cal A}_s}{M^\star}
  \;\;\;\;\;\;
  a_v\;=\; \frac{{\cal A}_v^\mu P^\star_\mu}{P^{\star\,2}}
  \;\;\;\;\;\;
  a_t\;=\; \frac{{\cal A}_t^{\mu\nu} T_{\mu\nu}}{T^2}
\;,\end{equation}
similar for $g_{s,v,t}\leftrightarrow G$ and  
$k_{s,v,t}\leftrightarrow\Gamma$. The scalar equations are
\begin{eqnarray}\label{ss3} \nonumber
P^{\star\,2}\,g_v - M^{\star\,2}\,g_s
  -2 \,T^2\,g_t&=&
1  - \pi^2
  \;\left( M^{\star\,2}\,k_s a_s+\Gamma_{v,\mu}\,{\cal A}_v^\mu
  +2 \Gamma_{t,\mu\nu}{\cal A}_t^{\mu\nu}
  \right)\\
P^{\star\,2}\,a_v - M^{\star\,2}\,a_s
  -2 \,T^2\,a_t&=&
  M^{\star\,2}\,k_s g_s+\Gamma_{v,\mu}\,G_v^\mu
  +2 \Gamma_{t,\mu\nu}G_t^{\mu\nu}
\;.\end{eqnarray}
The vector equations are contracted with $P^\star_\nu$ first:
\begin{eqnarray}\label{sv4}\nonumber
g_s - g_v
 &=&- \pi^2     \left( k_s\,a_v+k_v\,a_s \right)\\
a_s - a_v  &=&   k_s\,g_v+k_v\,g_s
\;,\end{eqnarray}
and then with $\Gamma_{v,\nu}$:
\begin{eqnarray}\label{sv5}\nonumber
  P^{\star\,2}\,k_v\,g_s - G_v^\nu\Gamma_{v,\nu}
 &=&- \pi^2
            \left( k_s\,{\cal A}_v^\nu\Gamma_{v,\nu}
                 + a_s\, \Gamma_v^\nu\Gamma_{v,\nu} \right)\\
  P^{\star\,2}\,k_v\,a_s - {\cal A}_v^\nu\Gamma_{v,\nu}
 &=&          k_s\,G_v^\nu\Gamma_{v,\nu}
                 + g_s\,\Gamma_v^\nu\Gamma_{v,\nu}
\;.\end{eqnarray}
The tensorial equations are contracted with
$P^{\star\,\nu} T^{\sigma\tau}$, leading to
\begin{eqnarray}\label{st3}\nonumber
g_t-g_v &=& -\pi^2\left(k_t\,a_v+k_v\,a_t\right)\\
a_t-a_v &=& k_t\,g_v+k_v\,g_t
\;,\end{eqnarray}
and with
$P^{\star\,\nu} \Gamma_t^{\sigma\tau}$:
\begin{eqnarray}\label{st4}\nonumber
G_{t,\sigma\tau}\Gamma_t^{\sigma\tau}
  - T^2\, g_v k_t &=&-\pi^2\left( a_v\,
  \Gamma_{t,\sigma\tau}\Gamma_t^{\sigma\tau}
  +k_v\,{\cal A}_{t,\sigma\tau}\Gamma_t^{\sigma\tau}\right)\\
{\cal A}_{t,\sigma\tau}\Gamma_t^{\sigma\tau}
  - T^2\, a_v k_t &=& g_v\,
  \Gamma_{t,\sigma\tau}\Gamma_t^{\sigma\tau}
  +k_v\,G_{t,\sigma\tau}\Gamma_t^{\sigma\tau}
\;,\end{eqnarray}
We can then use eqns. \fmref{ss3} and \fmref{st4} to eliminate
the products of ${\cal A}$ and $G$ with $\Gamma$.
The abbreviations used in the following are defined in eqns. \fmref{xqdef},
\fmref{db2}, and the final solution eqn. \fmref{sol5} is obtained through
\begin{eqnarray}\label{sol4}\nonumber
{\cal D}\,a_s\;=\;k_s {\cal X} + {\cal W}\;\;,\;\;&
{\cal D}\,a_v\;=\;-k_v {\cal X} + {\cal W}\;\;,\;\;&
{\cal D}\,a_t\;=\;k_t {\cal X} + 2\,{\cal W}\\
{\cal D}\,g_s\;=\;{\cal X} -k_s\pi^2 {\cal W}\;\;,\;\;&
{\cal D}\,g_v\;=\;{\cal X} +k_v\pi^2 {\cal W}\;\;,\;\;&
{\cal D}\,g_t\;=\;{\cal X} -k_t\pi^2 {\cal W}
\;.\end{eqnarray}
\clearpage
\section{Dissipative kernel of the transport equation}\label{dik}
In this section, the expression on the right side of the
quantum transport equation \fmref{t2} is evaluated to first order 
in the gradient expansion. First, we write it to infinite
order in $\Diamond$
\begin{equation}\label{t42} 
{\cal C}_{XP}=\exp(-{\mathrm i}\Diamond)\;
\widetilde{\Gamma}_{XP} \left( G_{XP}
               +   {\mathrm i}\pi{\cal A}_{XP}\right) 
\;.\end{equation}
The next step
consists of using \fmref{raff2} in connection with \fmref{gex}
(See remark in the text after eqn. \fmref{t4}). We remove the
gradient expansion for this intermediate step, 
\begin{eqnarray} \nonumber
{\cal C}_{XP}
  &=&2\pi \int\!\!d^4z\,\frac{d^4Q}{(2\pi)^4} \;\Theta(-z_0)
 \exp\left({\mathrm i}(P-Q)z +\frac{{\mathrm i}}{2}\,
   \partial_X^\Gamma\partial_P^A
   \right)\,\times\\ 
&&\;\;\;\;\;\;\;\;\;\;\;\;\;
   \widetilde{\Gamma}(X+z/2;P)\,{\mathrm i}{\cal A}(X;Q)  
\;.\end{eqnarray}
The three-dimensional integrations can be carried out,
with $X=(t,\vec{x})$ and $P=(p_0,\vec{p})$ one obtains
\begin{eqnarray} \nonumber
{\cal C}_{XP}
&=&{\mathrm i}\int\!\!d\tau\,dE\;
\Theta(-\tau)\,\exp\left(-{\mathrm i}\Diamond-{\mathrm i}\tau 
    E+\frac{{\mathrm i}}{2}\,
   \partial_{t}^\Gamma\partial_{p_0}^A
   \right)\,\times\\ 
&&\;\;\;\;\;\;\;\;\;\;\;\;\;
   \widetilde{\Gamma}(t+\tau/2,\vec{x};P)
   \;{\cal A}(X;p_0+E,\vec{p})
\:.\end{eqnarray}
The solutions for $G$ and ${\cal A}$ are known
only to first order in the gradient expansion, hence this expression
is also valid only in first order in $\Diamond$. Furthermore,
the additional derivative term in the exponent is eliminated
by shifting the argument of $\widetilde{\Gamma}$:
\begin{eqnarray}\nonumber
{\cal C}_{XP}&=&
{\mathrm i}\int\!\!d\tau\,dE\;
\Theta(-\tau)\,\exp(-{\mathrm i}\tau E)\,
   \left(1-{\mathrm i}\Diamond+\frac{{\mathrm i}}{2}
   \partial_{t}^\Gamma\partial_{p_0}^A
   \right)\,\times\\ 
&&\;\;\;\;\;\;\;\;\;\;\;\;\;
   \widetilde{\Gamma}(t+\tau/2,\vec{x};P)
   \;{\cal A}(X;p_0+E,\vec{p})\\ \nonumber
&=&{\mathrm i}\int\!\!d\tau\,dE\;
\Theta(-\tau)\,\exp(-{\mathrm i}\tau E)\,
 \left(\widetilde{\Gamma}(X;P)\;
 \vphantom{\int}{\cal A}(X;p_0+E,\vec{p})\right.\\
\nonumber
&&\;\;\;\;\;\;\;\;\;\;\;\;\;\;\;\;\;\;\;\;\;\;\;\;\;\;\left.
   -{\mathrm i}\Diamond\,\widetilde{\Gamma}(t+\tau/2,\vec{x};P)
   \vphantom{\int}\;{\cal A}(X;p_0+E,\vec{p})\right)
\;.\end{eqnarray}
In the first part, the $t$-integration can be carried out, 
\begin{eqnarray} \nonumber
{\cal C}_{XP}
&=&\int\!\!dE\;\frac{1}{p_0-E-{\mathrm i}\epsilon}\;
 \widetilde{\Gamma}(X;P)\;{\cal A}(X;E,\vec{p}\\
&+&
\int\!\!d\tau\,dE\;
\Theta(-\tau)\,\exp(-{\mathrm i}\tau E)\;
   \Diamond\,\widetilde{\Gamma}(t+\tau/2,\vec{x};P)
   \;{\cal A}(X;p_0+E,\vec{p})
\;.\end{eqnarray}
Together with \fmref{raff}, this is equivalent to eqn.
\fmref{t4}.


\clearpage
\begin{thebibliography}{99}
\bibitem{GCM93}{
   Gy.Wolf, W.Cassing and U.Mosel, Nucl.Phys. {\bf A 552} (1993) 549}
\bibitem{D84a}{
    P.Danielewicz,
    Ann.Phys. {\bf 152} (1984)  239 and 305}
\bibitem{RS86}{
    J.Rammer and H.Smith,
    Rev.Mod.Phys. {\bf 58} (1986) 323}
\bibitem{MD90}{
    S.Mrowczynski and P.Danielewicz,
    Nucl.Phys. {\bf B 342} (1990) 345}
\bibitem{M86}{
    S.Mrowczynski, Ann.Phys. {\bf 169} (1986) 48}
\bibitem{NBH93}{
    L.W.Neise, V.Bunakov and J.H\"ufner,\\
    {\em Beyond the Quasi Particle Approximation}\\
    {\em in Kinetic Theory for Heavy-Ion Collisions}\\
    U Heidelberg preprint, 1993}
\bibitem{WR93}{
    K.Wagner and P.G.Reinhard,\\
    Proc. of the Int. Workshop on Gross Properties of Nuclei\\
    and Nuclear Excitations, Hirschegg 1993;\\
    ed. H.Feldmeier (GSI Darmstadt, 1993) 320}
\bibitem{R94}{
    J.Rau\\
    {\em Pair production in the Quantum Boltzmann equation\/},\\
    Duke university preprint DUKE-TH-93-55 (hep-ph 9402256)}
\bibitem{HFN93}{
    M.Herrmann, B.Friman and W.N\"orenberg,\\
    GSI-Report 92-10  (1992) and Nucl.Phys. {\bf A560} (1993) 411}
\bibitem{BS75}{
    H.J.Borchers, R.N.Sen,
    Commun.Math.Phys.{\bf 21} (1975) 101}
\bibitem{NRT83}{
    H.Narnhofer, M.Requardt and W.Thirring,\\
    Commun.Math.Phys. {\bf 92} (1983) 247}
\bibitem{L88}{
    N.P.Landsman, Ann.Phys. {\bf 186 } (1988) 141}
\bibitem{SKM}{
    J.Schwinger,
    J.Math.Phys. {\bf 2} (1961) 407;\\[1mm]
    L.V.Keldysh, Zh.Exsp.Teor.Fiz. {\bf 47} (1964) 1515 and
    JETP {\bf 20} (1965) 1018}
\bibitem{NUY92}{
    K.Nakamura, H.Umezawa and Y.Yamanaka,
    Mod.Phys.Lett {\bf A7} (1992)  3583}
\bibitem{hu92}{
    P.A.Henning and H.Umezawa\\
    {\em Diagonalization of Propagators in Thermo Field Dynamics for}\\
    {\em Relativistic Quantum Fields}\\
     GSI Preprint 92-61 (1992), Nucl.Phys. {\bf B} in press}
\bibitem{Ubook}{
    H.Umezawa,\\
    { Advanced Field Theory: Micro, Macro and Thermal Physics}\\
    (American Institute of Physics, 1993)}
\bibitem{hu92a}{
    P.A.Henning and H.Umezawa,
    Phys.Lett. {\bf B303} (1993) 209}
\bibitem{h93trans}{
    P.A.Henning \\
    {\em TFD and Kinetic Coefficients of a Charged Boson Gas},\\
    GSI-Preprint 93-45 (1993), Nucl.Phys. {\bf A} in press}
\bibitem{E93}{
    T.S.Evans, Phys.Rev. {\bf D47} (1993) R4196}
\bibitem{MH93}{
    S.Mrowczynski and U.Heinz,  U Regensburg preprint (1992)}
\bibitem{W67}{
   A.S.Wightman, {\em in:\/} 
   { Carg\`ese Lectures in Theoretical
   Physics:\\  High Energy Electromagnetic Interactions and
   Field Theory,}\\  ed. M. L\'evy
   (Gordon \& Breach, New York 1967)}
\bibitem{PW88a}{
    P.Poschenrieder and M.K.Weigel,
    Phys.Rev. {\bf C 38} (1988) 471}
\bibitem{h91spec}{
    P.A.Henning, Phys.Lett {\bf B272} (1991) 186
    and Z.Physik  {\bf A345} (1993) 227}
\bibitem{BM90}{
    W.Botermans and R.Malfliet,
    Phys.Rep. {\bf 198} (1990) 115}
\bibitem{EW92}{
    M.A.van Eijck and Ch.G. van Weert,
    Phys.Lett. {\bf B278} (1992) 305}
\bibitem{E92}{
  T.S.Evans, Nucl.Phys. {\bf B374} (1992) 340}    
\bibitem{habil}{
    P.A.Henning,\\
    {\em TFD for Quantum Fields with Continuous Mass Spectrum\/}\\
    Habilitationsschrift (October 1993)}
\bibitem{h92fock}{
    P.A.Henning,
    Nucl.Phys. {\bf A 546} (1992) 653}
\bibitem{W60}{
    S.Weinberg, Phys.Rev. {\bf 118} (1960) 848}
\bibitem{KB62}{
  L.P.Kadanoff and G.Baym,
  { Quantum Statistical Mechanics}
  (Benjamin, Reading 1962)}
\bibitem{L90}{
    R.L. Liboff, { Kinetic Theory}
    (Prentice-Hall, Englewood Cliffs, NJ 1990)}
\bibitem{KMS}{
    R. Kubo,
    J.Phys.Soc. Japan {\bf 12} (1957) 570;\\[1mm]
    C.Martin and J.Schwinger,
    Phys.Rev. {\bf 115} (1959) 1342}\end{thebibliography}
\end{document}